\documentstyle[11pt]{article}
\begin{document}

\begin{center}
{\bf ESTIMATIVE OF INELASTICITY COEFFICIENT \\ FROM EMULSION CHAMBER DATA}

\vspace{.3 true cm}

{H M PORTELLA}

\vspace{.05 true cm}

{\it Instituto de F\'{\i}sica, Universidade Federal Fluminense, Av. Litor\^anea s/n, \\ Gragoat\'a, 24210-340, Niter\'oi, RJ, Brazil}

\vspace{.1 true cm}

{L C S de OLIVEIRA and C E C LIMA}

\vspace{.05 true cm}

{\it Centro Brasileiro de Pesquisas F\'{\i}sicas, Rua Dr. Xavier Sigaud 150, \\ Urca, 22290-180, Rio de Janeiro, RJ, Brazil}
\end{center}

\vspace{.2 true cm}

\abstract{Atmospheric diffusion of high energy cosmic rays is studied analytically and the obtained integral electromagnetic fluxes are compared with the data measured by emulsion chamber detectors at mountain altitudes. We find a good consistency between them when the average nucleon inelasticity coefficient is varying between 0.50 and 0.65.} 

\section{Introduction}

\noindent

The inelasticity is a global parameter of hadronic interactions and it is a crucial attribute in the calculation of the fluxes in cosmic ray cascades. It is defined as the fraction of energy giving up by the leading hadron in a collision induced by an incident hadron on a target nucleon or nucleus. The inelasticity, $K_{\mbox {\tiny $N$}}$, in a nucleon-induced interaction is related to the spectrum-weighted moments evaluated at $\gamma = 1$ (where $\gamma$ is the power index of primary spectrum) as 
$$1 - K_{\mbox {\tiny $N$}} = Z_{{\mbox {\tiny $NN$}}}(\gamma = 1) = \int_0^1 \eta \mbox{\hspace{.1 true cm}} f_{{\mbox {\tiny $NN$}}}(\eta) \mbox{\hspace{.1 true cm}} d\eta \mbox{\hspace{.2 true cm} .}$$
\par
Several authors have suggested that the average inelasticity-coefficient is an increasing function of the energy \cite{ref1,ref2}, whereas others proposed that it is a decreasing function \cite{ref3,ref4,ref5}. The behaviour of high energy cosmic rays, which reflects the nuclear interactions in the energy region $1~\sim 100$~TeV, indicates a constant value of the mean inelasticity equal to 0.50 \cite{ref6}. A decreasing with energy of this parameter points out a strongly suppression of the pion production in this energy region and does not agree with cosmic ray observations.

The objective of this paper is to estimate the average inelasticity coefficient through the comparison of our calculations with the integral electromagnetic fluxes measured with large Emulsion Chamber Experiments at mountain altitudes. The main purpose is to describe in qualitative detail single events, {\it i.e.}, shower cores in their early stage of development observed at mountain altitudes.

Our calculation is performed by solving analitically the cosmic ray diffusion equations for  electromagnetic showers in earth's atmosphere.

In these calculations we adopt to the particle production an approximated Feynman scaling in the fragmentation region. We have used the data on proton targets and on nuclear targets \cite{ref7} to obtain the pion spectrum weighted moments. Although exist several functional forms to fit the behaviour of rising cross-section we assume the following one 
$$\sigma = \sigma_{\mbox {\tiny 0}} \mbox{\hspace{.1 true cm}} (E/\epsilon)^\alpha \mbox{\hspace{.2 true cm} .}$$
\par
So, beginning with the measured all-particle primary spectrum we diffuse the particle showers down to the detection depths and we investigate our numerical results as a function of the adopted average inelasticity coefficient.

\section{Hadronic and electromagnetic cascades in the Earth's atmosphere}

\noindent

The differential hadronic fluxes, $H(t,E)$ at a depth $t$ in the energy range $E$ and $E + dE$ are obtained integrating the nucleon and pion diffusion equations using the semigroup theory. We assume that only the pions are generated in the multiparticle production neglecting the particles with small fractions such as kaons, heavy mesons, etc.. In this case the solutions are \cite{ref8};
\begin{equation}
N(t,E) = \sum_{n=0}^\infty \frac{(-1)^n}{n!} \left( \frac{t}{\lambda_{\mbox {\tiny $N$}}(E)} \right)^n \prod_{i=1}^n \left( 1 - Z_{{\mbox {\tiny $NN$}}}(\gamma,\alpha,n) \right) N(0,E)
\label{N-1}
\end{equation}
\noindent
with
\begin{equation}
Z_{{\mbox {\tiny $NN$}}}(\gamma,\alpha) = \int_0^1 f_{{\mbox {\tiny $NN$}}}(\eta) \mbox{\hspace{.1 true cm}} \eta^{\gamma - i \alpha } \mbox{\hspace{.1 true cm}} d\eta
\end{equation}
\noindent
where $\eta$ is the elasticity coefficient of the nucleons in the atmosphere that is distibuted according to $f_{{\mbox {\tiny $NN$}}}(\eta)$. 

The pionic fluxes are 
$$\Pi(t,E) = \sum_{m=0}^\infty \sum_{n=0}^\infty \int_0^t dz \mbox{\hspace{.1 true cm}} \frac{(-1)^m (-1)^n}{m! \mbox{\hspace{.1 true cm}} n!} \left( \frac{t-z}{\lambda_\pi(E)}\right)^m \left( \frac{z}{\lambda_{\mbox {\tiny $N$}}(E)}\right)^n Z_{{\mbox {\tiny $N$}}\pi}(\gamma,\alpha,n) \cdot$$
\begin{equation}
\cdot \prod_{i=1}^n \left( 1 - Z_{{\mbox {\tiny $NN$}}}(\gamma,\alpha,n) \right) \prod_{j=1}^m \left( 1 - Z_{\pi\pi}(\gamma,\alpha,m,n) \right) \frac{N(0,E)}{\lambda_{\mbox {\tiny $N$}}(E)}
\label{pi-1}
\end{equation}
\noindent
with
\begin{equation}
Z_{\pi\pi}(\gamma,\alpha,m,n) = \int_0^1 f_{\pi\pi}(x) \mbox{\hspace{.1 true cm}} x^{\gamma - \alpha (j + n + 1)} \mbox{\hspace{.1 true cm}} dx
\end{equation}
\noindent
and
\begin{equation}
Z_{{\mbox {\tiny $N$}}\pi}(\gamma,\alpha,n) = \int_0^1 f_{{\mbox {\tiny $N$}}\pi}(x) \mbox{\hspace{.1 true cm}} x^{\gamma - \alpha (n + 1)} \mbox{\hspace{.1 true cm}} dx \mbox{\hspace{.2 true cm} .}
\end{equation}
\par
The $f_{{\mbox {\tiny $N$}}\pi}(x)$ and $f_{\pi\pi}(x)$ are the energy spectra of the pions produced in the nucleon-air and in the pion-air interactions respectively, and $x$ is the Feynman variable $(x = 2p_L/\sqrt{S})$.

The solutions (\ref{N-1}) and (\ref{pi-1}) are subjected to the boundary conditions 
\begin{equation}
N(0,E) = N_{\mbox {\tiny 0}} \mbox{\hspace{.1 true cm}} E^{-(\gamma + 1)}
\label{bound-N}
\end{equation}
\noindent
and
\begin{equation}
\Pi(0,E) = 0
\label{bound-pi}
\end{equation}
\noindent
that represents the nucleon and pion fluxes at the top of atmosphere.

The above solutions are obtained considering the interaction mean-free paths in the power form,  $\lambda_{\mbox{ \tiny $H$}}(E) = \lambda_{{\mbox {\tiny 0}}_{\mbox {\tiny $H$}}} \mbox{\hspace{.1 true cm}} E^{-\alpha}$ (${\mbox{\tiny $H$}} = {\mbox {\tiny $N$}}$ or $\pi$).

Using naive approximations \cite{ref9} we can write the solutions (\ref{N-1}) and (\ref{pi-1}) in the following forms
\begin{equation}
N(t,E) = N(0,E) \mbox{\hspace{.1 true cm}} e^{-t \mbox{\hspace{.05 true cm}} \omega_{\mbox {\tiny $N$}}(E)}
\label{N-2}
\end{equation}
\noindent
and
\begin{equation}
\Pi(t,E) = N(0,E) \frac{\omega_{{\mbox {\tiny $N$}}\pi}(\gamma,E)}{\omega_{\mbox {\tiny $N$}}(E) - \omega_\pi(E)} \left( e^{-t \mbox{\hspace{.05 true cm}} \omega_\pi(E)} - e^{-t \mbox{\hspace{.05 true cm}} \omega_{\mbox {\tiny $N$}}(E)} \right) \mbox{\hspace{.2 true cm} .}
\label{pi-2}
\end{equation}
\par
The $\omega_{\mbox {\tiny $N$}}(E)$, $\omega_\pi(E)$ and the $\omega_{{\mbox {\tiny $N$}}\pi}(\gamma,E)$ functions are described in the appendix A.

The production rate of $\gamma$-rays originated from the $\pi^0$ decay is given by \cite{ref10}
\begin{equation}
P_\gamma(z,E_\gamma) = 2 \int_{E_\gamma}^\infty \frac{dE'}{2 E'} \left( P^{{\mbox {\tiny $N$}} - air}_{\pi^\pm} + P^{\pi^\pm - air}_{\pi^\pm}\right)
\label{P-gamma}
\end{equation}
\noindent
where $P^{{\mbox {\tiny $N$}} - air}_{\pi^\pm}$ and $P^{\pi^\pm - air}_{\pi^\pm}$ are the production rate of the charged pions from the nucleon-air and pion-air interactions and are represented by
\begin{equation}
P^{{\mbox {\tiny $H$}}-air}_{\pi^\pm} = \int_0^1 \frac{H(z,E_0)}{\lambda_{\mbox {\tiny $H$}}(E_0)} \mbox{\hspace{.1 true cm}} f_{{\mbox {\tiny $H$}}\pi}(x) \mbox{\hspace{.1 true cm}} \frac{d x}{x} \mbox{\hspace{.5 true cm} ; \hspace{.5 true cm} {\tiny $H$} = {\tiny $N$} or $\pi$ \hspace{.2 true cm} .}
\end{equation}
\par
In the expression (\ref{P-gamma}) we have used that the secondary pions are formed with equal multiplicity, so the number of $\pi^0$ is half of the number of charged pions.

The differential flux of the electromagnetic component can be written as
\begin{equation}
F_{\gamma,e}(t,E) = \int_0^t dz \int_E^\infty dE_\gamma \mbox{\hspace{.1 true cm}} P_\gamma(z,E_\gamma) \mbox{\hspace{.1 true cm}} (e + \gamma)(t-z,E_\gamma,E)
\label{F-gamma}
\end{equation}
\noindent
where $(e + \gamma)(t-z,E_\gamma,E)$ are the photons, electrons and positrons produced by a single photon. They are computed using the approximation {\bf A} \cite{ref11} and are written in the following form
\begin{equation}
(e + \gamma)(t-z,E_\gamma,E) = \frac{1}{2\pi i} \int_{\mbox {\tiny $C$}} \left( \frac{E_\gamma}{E} \right)^s \frac{1}{E} \left( N_1(s) e^{\lambda_1(s)(t-z)/X_0} + N_2(s) e^{\lambda_2(s)(t-z)/X_0} \right)
\label{e-gamma} \mbox{\hspace{.2 true cm} .}
\end{equation}
\par
The functions $N_i(s)$, $\lambda_i(s)$ and $X_0$ are the parameters with standardized definitions in cascade theory \cite{ref11}. These functions and the evaluation of the electromagnetic fluxes are presented in the appendix B.

In order to compare our calculations for electromagnetic showers with the emulsion chamber data measured at mountain altitudes we need to obtain the integral fluxes of this component. They are, 
%\begin{equation}
%H(t,\geq E_h^{(\gamma)}) = \int_{E_h^{(\gamma)}}^\infty H(t,E_h^{(\gamma)}) \mbox{\hspace{.1 %true cm}} dE_h^{(\gamma)}
%\label{H-gamma}
%\end{equation}
%\noindent
%and
\begin{equation}
F_{\gamma,e}(t,\geq E) = \int_E^\infty F_{\gamma,e}(t,E) \mbox{\hspace{.1 true cm}} dE
\label{F-e-gamma}
\end{equation}

\section{Comparison with experimental data}

\noindent

Emulsion Chamber Experiments at mountain altitudes are very simple detectors constituted of multiple sandwich of material layers (Pb, Fe, etc.) and photo-sensitive layers (X-ray films, nuclear emulsions plates, etc.). They can detect electromagnetic showers of high energy ($E_{e,\gamma} \geq$ 1 TeV) and due to its powerful spatial resolution and energy determination they can study in details the cosmic-ray interactions.

The experimental data are taken from J. R. Ren {\it et al.} \cite{ref19} for Mt. Kanbala; C. M. G. Lattes {\it et al.} \cite{ref13} for Mt. Chacaltaya and M. Amenomori {\it et al.} \cite{ref21} for Mt. Fuji.

The experimental fluxes at Mt.~Chacaltaya are revised according to the energy determination introduced by M.~Okamoto and T.~Shibata \cite{ref22} where the LPM and spacing effects were taken into account. This revision followed the same procedure made by N.M.~Amato and N.~Arata \cite{ref20} for determination of total hadron fluxes.

In order to make a comparison with electromagnetic fluxes measured with these detectors it is necessary to check the main assumptions made in our calculations;

\noindent
a) Primary cosmic-ray radiation

\vspace{.2 true cm}

The energy spectra obtained by Emulsion Chamber Experiments at mountain altitude are in the region of 3 $\sim$ 50 TeV for $(e,\gamma)$ component and 3 $\sim$ 40 TeV for hadronic shower. So, we need to include in our calculations the primary spectrum up to knee ($\approx 10^{15}$ eV). In the present work we assume the primary cosmic-ray radiation in integral form reported by 
J.~Bellandi Filho {\it et al.} \cite{ref14} from JACEE data \cite{ref15}.
\begin{equation}
N_1(0,\geq E) = 1.12 \cdot 10^{-6} (E/5 \mbox{TeV})^{-1.7} \mbox{(/cm$^2$.s.st)}
\end{equation}
\noindent
where $E$(TeV/nucleus).

Nucleonic intensity at $t=0$, equivalent to primary cosmic-ray radiation is given by 
\begin{equation}
N(0,\geq E) = A^{1-\gamma}N_1(0,\geq E)
\end{equation}
\noindent
where $A$ is the average mass number of cosmic ray nuclei; we adopt $A = 14.0$ which is the value in low energy region $(E$/part~$<~10^{14}$~eV).

\vspace{.2 true cm}

\noindent
b) Collision mean free paths

\vspace{.2 true cm}

We use to fit the behaviour of rising with energy of the p-air inelastic cross-section the following form \cite{ref7,ref14,ref16,ref17}, 
\begin{equation}
\sigma_N(E) = \sigma_{N_0} (E/\mbox{TeV})^\alpha
\end{equation}
\noindent
with $\sigma_{N_0} = 300$ mb and $\alpha = 0.06$ \cite{ref7,ref14}.

So, the collision mean free path of the nucleon in the earth's atmosphere is expressed by
\begin{equation}
\lambda_N(E) = \lambda_{N_0} (E/\mbox{TeV})^{-\alpha}
\end{equation}
\noindent
with $\lambda_{N_0}= 80$ g/cm$^2$ \cite{ref7,ref14}.

For the pion mean free paths we assume that $\lambda_{\pi_0}/\lambda_{N_0} \approx 1.4$ and that they have the same energy dependence, like in the nucleon case \cite{ref14}. Thus, we have
\begin{equation}
\lambda_\pi(E) = \lambda_{\pi_0} (E/\mbox{TeV})^{-\alpha}
\end{equation}
\noindent
with $\lambda_{\pi_0}= 112$ g/cm$^2$. 

\vspace{.2 true cm}

\noindent
c) Secondary pions:

\vspace{.2 true cm}

The functions $f_{{\mbox {\tiny $N$}}\pi}(x)$ and $f_{\pi\pi}(x)$ are the energy distribution of secondary nucleons and pions for nucleon-air and pion-air interactions, respectively. We have used in this calculation the data on proton targets and on nuclear targets \cite{ref7} to obtain these functions and they are written as 
\begin{equation}
f_{{\mbox {\tiny $N$}}\pi}(x) = 2.08 \mbox{\hspace{.1 true cm}} \frac{(1-x)}{x} \mbox{\hspace{.1 true cm}} e^{-5 x}
\label{fNpi}
\end{equation}
\noindent
and
\begin{equation}
f_{\pi\pi}(x) = \frac{2.6}{x}\left( 1 + \frac{x}{0.45}\right)^{-3} + \frac{0.32}{x} \mbox{\hspace{.1 true cm}} e^{2(x-1)}
\label{fpipi}
\end{equation}
\noindent
these functions are for two charged states of pions and the contribution of leading pions are taken into account in the last one. 

Figures 1-3 show the comparison of our solution with the integral electromagnetic fluxes measured at Mt. Kanbala (520 g/cm$^2$), Mt. Chacaltaya (540 g/cm$^2$) and Mt. Fuji (650 g/cm$^2$), respectively. Two curves are also drawn in the figures for $< K_N >$ = 0.50 and 0.65, taking into account the above mentionated itens. 

\section{Conclusions}

\noindent

We have solved the diffusion equations of cosmic-ray hadrons analytically and after this we derived the integral electromagnetic fluxes using the approximation {\bf A} of the electromagnetic cascade theory.

These fluxes at mountain atmospheric depths decrease when we include in our calculation the rising of the cross-section and the decreasing of the mean nucleon elasticity.

We conclude that the naive assumptions for the secondary production based on approximate scaling in the fragmentation region is possible to explain experimental data on integral electromagnetic fluxes detected in large emulsion chamber experiments.

This scenario requires a mean nucleon inelasticity coefficient varying from 0.50 to 0.65. At low energy (few TeV) the agreement between experimental data and the calculated fluxes shows that 
$< K_N > \approx 0.50$, that is the adjust indicated by former workers in the cosmic ray literature. For $E > 10$ TeV we see from the figure that $< K_N > \approx 0.65$. Therefore, with this simple analysis we see that the nucleon inelasticity coefficient appears to show a tendency of rising with energy according to Mini-jet Models \cite{ref17} and Quark-Gluon String Models \cite{ref2}. These integral fluxes are somewhat affected when the coefficient $\alpha$ changes in the interval 0.06 to 0.10, which is the appropriated values to explain the experimental data for mountain emulsion chambers experiments. In the latter the fit between our calculation and the experimental data is obtained for $<~K_N~>$ changing between approximately 0.45 and 0.60. 

These results are in agreement with that of L. Jones \cite{ref18} based in an analysis on inclusive reactions data from accelerator, with analytical calculations on nucleon fluxes at sea level \cite{ref9} and with hadron and electromagnetic fluxes at mountain altitudes \cite{ref8,ref10}.

\section*{Appendix A}

\noindent
{\bf Hadronic cascades}

\vspace{.2 true cm}

The functions $\omega_{\mbox {\tiny $N$}}(E)$, $\omega_\pi(E)$ and $Z_{{\mbox {\tiny $N$}}\pi}(\gamma,E)$ in the hadronic fluxes (\ref{N-2}) and (\ref{pi-2}) are \cite{ref9}
\begin{equation}
\omega_{\mbox {\tiny $N$}}(E) = \omega_{\mbox {\tiny $N$}} \mbox{\hspace{.1 true cm}} E^{0.06} = \frac{1}{\lambda_{\mbox {\tiny $N$}}(E)}\left( 1 - \frac{1}{2}Z_{{\mbox {\tiny $NN$}}}(\gamma) - \frac{1}{2}Z_{{\mbox {\tiny $NN$}}}(\gamma,\alpha) \right)
\end{equation}
\begin{equation}
\omega_\pi(E) = \omega_\pi \mbox{\hspace{.1 true cm}} E^{0.06} = \frac{1}{\lambda_\pi(E)}\left( 1 + \frac{1}{2}Z_{\pi\pi}(\gamma) - \frac{3}{2}Z_{\pi\pi}(\gamma,\alpha) \right)
\end{equation}
\noindent
and
\begin{equation}
\omega_{{\mbox {\tiny $N$}}\pi}(E) = \frac{Z_{{\mbox {\tiny $N\pi$}}}(\gamma,\alpha)}{\lambda_{\mbox {\tiny $N$}}(E)} = \frac{1}{\lambda_{\mbox {\tiny $N$}}(E)} \int_0^1 x^{\gamma - \alpha} \mbox{\hspace{.1 true cm}} f_{{\mbox {\tiny $N$}}\pi}(x) \mbox{\hspace{.1 true cm}} dx
\end{equation}
\noindent
where
\begin{equation}
Z_{{\mbox {\tiny $NN$}}}(\gamma,\alpha) = \int_0^1 \eta^{\gamma - \alpha} \mbox{\hspace{.1 true cm}} f_{{\mbox {\tiny $NN$}}}(\eta) \mbox{\hspace{.1 true cm}} d\eta
\end{equation}
\noindent
and
\begin{equation}
Z_{\pi\pi}(\gamma,\alpha) = \int_0^1 x^{\gamma - \alpha} \mbox{\hspace{.1 true cm}} f_{\pi\pi}(x) \mbox{\hspace{.1 true cm}} dx \mbox{\hspace{.2 true cm} .}
\end{equation}
\par
The functions $f_{{\mbox {\tiny $N$}}\pi}(x)$ and $f_{\pi\pi}(x)$ are the energy distribution of secondary nucleons and pions for nucleon-air and pion-air interactions, respectively (see item c of section 3). 

The nucleon elasticity distribution is assumed to be \cite{ref9}
\begin{equation}
f_{{\mbox {\tiny $NN$}}}(\eta) = (1 + \beta)(1-\eta)^\beta
\end{equation}
\noindent
where $\eta~=~E/E'$ and $\beta$ fulfills a consistency relation between the mean elasticity and average inelasticity, so that $<~N~>~+~<~K~>~=~1$, for nucleons. When $\beta~=~0$ we have the nucleon uniform distribution of elasticity ($<~\eta~>~=~0.5$).

\section*{Appendix B}

\noindent
{\bf Electromagnetic cascades}

\vspace{.2 true cm}

The production rate of $\gamma$-rays produced from $\pi^0$ decay expression (\ref{P-gamma}) can be written as
$$P_\gamma(z,E_\gamma) = \frac{N_0 \mbox{\hspace{.1 true cm}} E_\gamma^{-(\gamma + 1 - \alpha + z \mbox{\hspace{.05 true cm}} \alpha \mbox{\hspace{.05 true cm}} \omega_{\mbox {\tiny $N$}})} e^{-z \mbox{\hspace{.05 true cm}} \omega_{\mbox {\tiny $N$}}}}{\gamma + 1 - \alpha + z \mbox{\hspace{.1 true cm}} \alpha \mbox{\hspace{.1 true cm}} \omega_{\mbox {\tiny $N$}}} \left( \frac{Z_{{\mbox {\tiny $N$}}\pi}(\gamma - \alpha + z \mbox{\hspace{.1 true cm}} \alpha \mbox{\hspace{.1 true cm}} \omega_{\mbox {\tiny $N$}} )}{\lambda_{{\mbox {\tiny $N$}}_{\mbox {\tiny 0}}}} - \frac{Z_{\pi\pi}(\gamma - \alpha + z \mbox{\hspace{.1 true cm}} \alpha \mbox{\hspace{.1 true cm}} \omega_{\mbox {\tiny $N$}} )}{\lambda_{\pi^0}}\right. $$
\begin{equation}
\left. \frac{\omega_{{\mbox {\tiny $N$}}\pi}(E)}{\omega_{\mbox {\tiny $N$}}(E) - \omega_\pi(E)}\right) +
\frac{N_0 E^{-(\gamma + 1 - \alpha + z \mbox{\hspace{.05 true cm}} \alpha \mbox{\hspace{.05 true cm}} \omega_\pi)}}{(\gamma + 1 - \alpha + z \mbox{\hspace{.1 true cm}} \alpha \mbox{\hspace{.1 true cm}} \omega_\pi )} \frac{\omega_{{\mbox {\tiny $N$}}\pi}(E) e^{-z \mbox{\hspace{.05 true cm}} \omega_\pi}}{(\omega_{\mbox {\tiny $N$}}(E) - \omega_\pi(E))}\frac{Z_{\pi\pi}(\gamma - \alpha + z \mbox{\hspace{.1 true cm}} \alpha \mbox{\hspace{.1 true cm}} \omega_\pi)}{\lambda_{\pi^0}} \mbox{\hspace{.2 true cm} .}
\label{B1}
\end{equation}
\par
The Z-factors $Z_{{\mbox {\tiny $N$}}\pi}$ and $Z_{\pi\pi}$ is the last expression are 
\begin{equation}
Z_{{\mbox {\tiny $N$}}\pi}(\gamma - \alpha + z \mbox{\hspace{.1 true cm}} \alpha \mbox{\hspace{.1 true cm}} \omega_{\mbox {\tiny $N$}}) = \int_0^1 x^{\gamma - \alpha + z \mbox{\hspace{.05 true cm}} \alpha \mbox{\hspace{.05 true cm}} \omega_{\mbox {\tiny $N$}}} \mbox{\hspace{.1 true cm}} f_{{\mbox {\tiny $N$}}\pi}(x) dx
\end{equation}
\noindent
and
\begin{equation}
Z_{\pi\pi}(\gamma - \alpha + z \mbox{\hspace{.1 true cm}} \alpha \mbox{\hspace{.1 true cm}} \omega_i) = \int_0^1 x^{\gamma - \alpha + z \mbox{\hspace{.05 true cm}} \alpha \mbox{\hspace{.05 true cm}} \omega_i} \mbox{\hspace{.1 true cm}} f_{\pi\pi}(x) \mbox{\hspace{.1 true cm}} dx \mbox{\hspace{.3 true cm} $i = {\mbox {\tiny $N$}}$ or $\pi$ .}
\end{equation}
\par
The expression (\ref{B1}) was obtained considering the expansion 
\begin{equation}
\frac{1}{\lambda_{\mbox {\tiny H}}(E)} = \frac{1 + \alpha \mbox{\hspace{.1 true cm}} ln(E/E_0)}{\lambda_{0_{\mbox {\tiny H}}}} \mbox{\hspace{.5 true cm} ; \hspace{.5 true cm} $H$ = {\mbox {\tiny $N$}} or $\pi$}
\end{equation}
\par
The differential flux of the electromagnetic component (\ref{F-gamma}) is obtained from this production rate and evaluated at the poles $s_1 = \gamma - \alpha + z \mbox{\hspace{.1 true cm}} \alpha \mbox{\hspace{.1 true cm}} \omega_{\mbox {\tiny $N$}}$ and $s_2 = \gamma - \alpha + z \mbox{\hspace{.1 true cm}} \alpha \mbox{\hspace{.1 true cm}} \omega_\pi$.

The integral flux is put in the form 
\begin{equation}
F_{\gamma,e}(t,\geq E) = F_{{\gamma}_1,e}(t,\geq E) + F_{{\gamma}_2,e}(t,\geq E)
\end{equation}
\noindent
with
\begin{equation}
F_{{\gamma}_1,e}(t,\geq E) = \int_0^t dz \mbox{\hspace{.1 true cm}} \frac{N_0 \mbox{\hspace{.1 true cm}} E^{-s_1} \mbox{\hspace{.1 true cm}} e^{-z \mbox{\hspace{.05 true cm}} \omega_{\mbox {\tiny $N$}}}}{s_1 (s_1 + 1)}\left( N_1(s_1)e^{\lambda_1(s_1)(t-z)/X_0} + N_2(s_1)e^{\lambda_2(s_1)(t-z)/X_0} \right) C_1(s_1)
\end{equation}
\noindent
and
\begin{equation}
F_{{\gamma}_2,e}(t,\geq E) = \int_0^t dz \mbox{\hspace{.1 true cm}} \frac{N_0 \mbox{\hspace{.1 true cm}} E^{-s_2} \mbox{\hspace{.1 true cm}} e^{-z \mbox{\hspace{.05 true cm}} \omega_\pi}}{s_2 (s_2 + 1)}\left( N_1(s_2)e^{\lambda_1(s_2)(t-z)/X_0} + N_2(s_2)e^{\lambda_2(s_2)(t-z)/X_0} \right) C_2(s_2)
\end{equation}
\noindent
where
\begin{equation}
C_1(s_1) = \frac{Z_{{\mbox {\tiny $N$}}\pi}(s_1)}{\lambda_{{\mbox {\tiny 0}}_{\mbox {\tiny $N$}}}} - \frac{\omega_{{\mbox {\tiny $N$}}\pi}(E)}{(\omega_{\mbox {\tiny $N$}}(E) - \omega_\pi(E))} \frac{Z_{\pi\pi}(s_1)}{\lambda_{0_\pi}}
\end{equation}
\noindent
and
\begin{equation}
C_2(s_2) = \frac{\omega_{{\mbox {\tiny $N$}}\pi}(E)}{(\omega_{\mbox {\tiny $N$}}(E) - \omega_\pi(E))} \frac{Z_{\pi\pi}(s_2)}{\lambda_{0_\pi}}
\end{equation}
\noindent
with
\begin{equation}
Z_{{\mbox {\tiny $N$}}\pi}(s_i) = \int_0^1 x^{s_i} \mbox{\hspace{.1 true cm}} f_{{\mbox {\tiny $N$}}\pi}(x) \mbox{\hspace{.1 true cm}} dx
\end{equation}
\noindent
and
\begin{equation}
Z_{\pi\pi}(s_i) = \int_0^1 x^{s_i} \mbox{\hspace{.1 true cm}} f_{\pi\pi}(x) \mbox{\hspace{.1 true cm}} dx \mbox{\hspace{.2 true cm} .}
\end{equation}
\par
The functions $N_i$ and $\lambda_i$ are the usual parameters with definitions in cascade theory and can be obtained in the reference \cite{ref11}. $X_0$ is the radiation length in the atmosphere ($X_0~=~37.1$~g/cm$^2$).

\newpage

\begin{center}
{\Large {\bf Figure Captions}}
\end{center}

\vspace{1.5 true cm}

\noindent
Figure 1 - Integral electromagnetic spectrum at Mt. Kanbala (520 g/cm$^2$). The full lines represent the calculated flux for $< K_N > = 0.65$ and the dashed lines for $< K_N > = 0.50$. All lines are for $\alpha = 0.06$. The experimental data are taken from J. R. Ren {\it et al.} \cite{ref19}.

\vspace{1.0 true cm}

\noindent
Figure 2 - Integral electromagnetic spectrum at Mt. Chacaltaya (540 g/cm$^2$). The full lines represent the calculated flux for $< K_N > = 0.65$ and the dashed lines for $< K_N > = 0.50$. All lines are for $\alpha = 0.06$. The experimental data are taken from C. M. G. Lattes {\it et al.}(black dots) \cite{ref13}.

\vspace{1.0 true cm}

\noindent
Figure 3 - Integral electromagnetic spectrum at Mt. Fuji (650 g/cm$^2$). The full lines represent the calculated flux for $< K_N > = 0.65$ and the dashed lines for $< K_N > = 0.50$. All lines are for $\alpha = 0.06$. The experimental data are taken from M Amenomori {\it et al.} \cite{ref21}. 

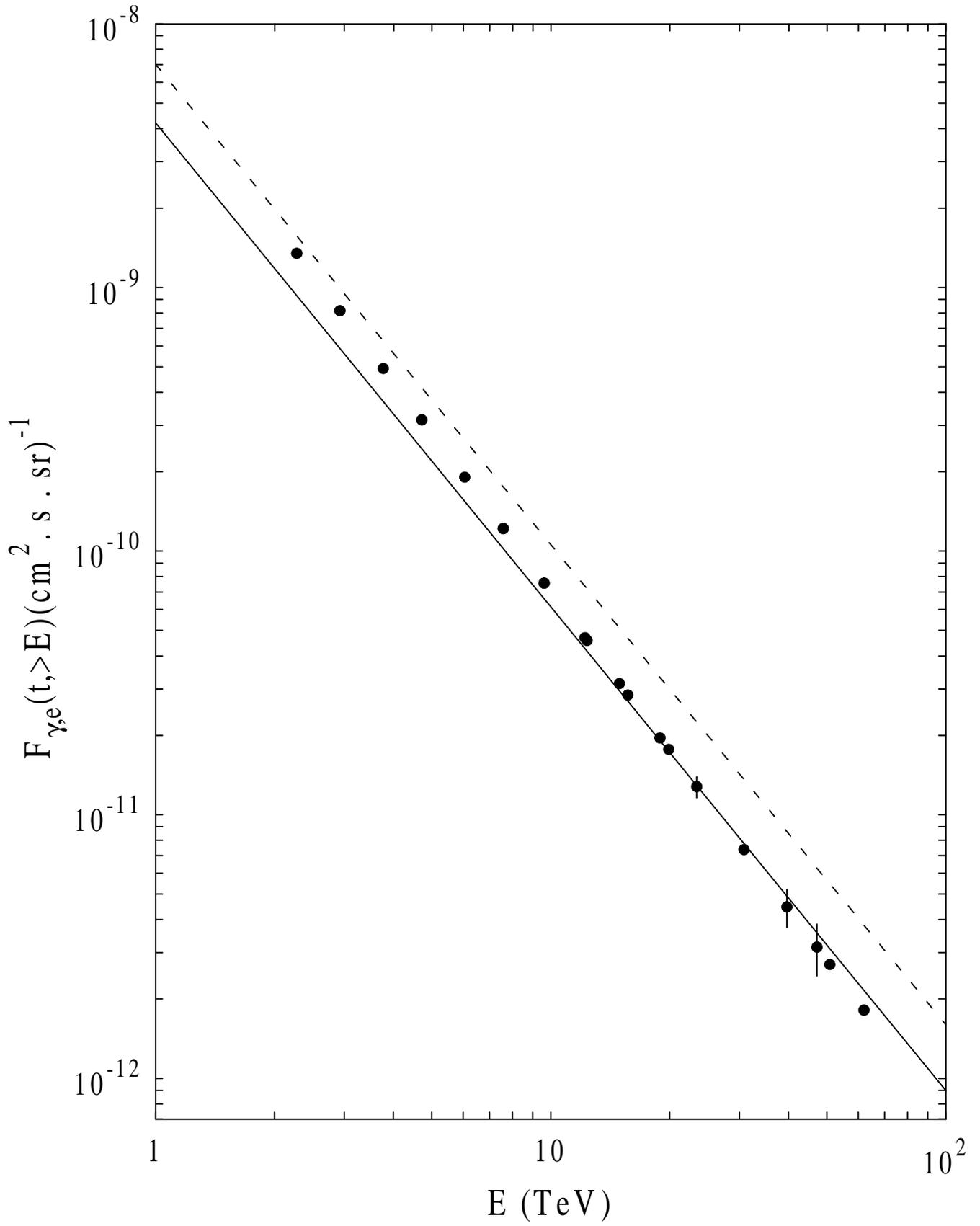
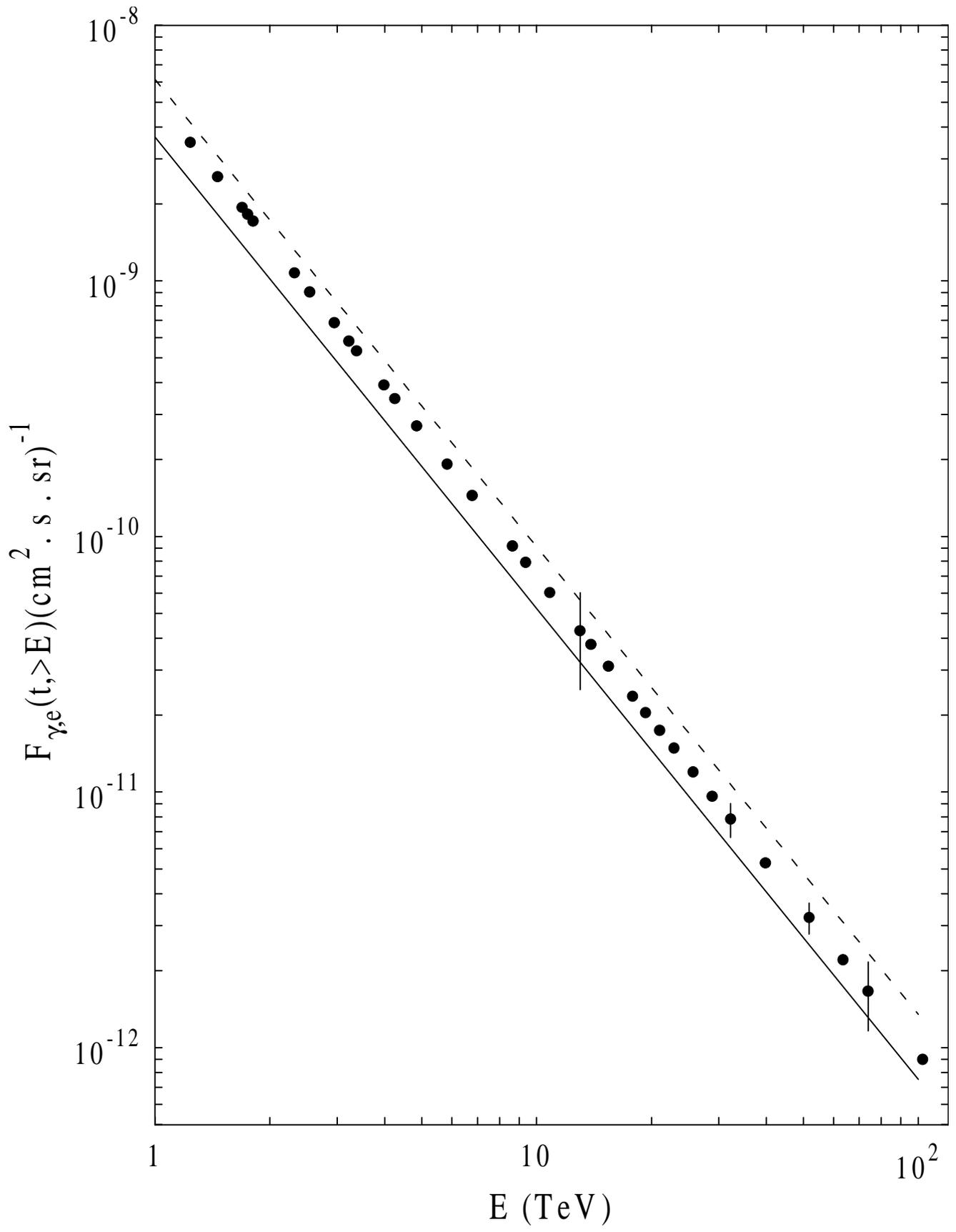
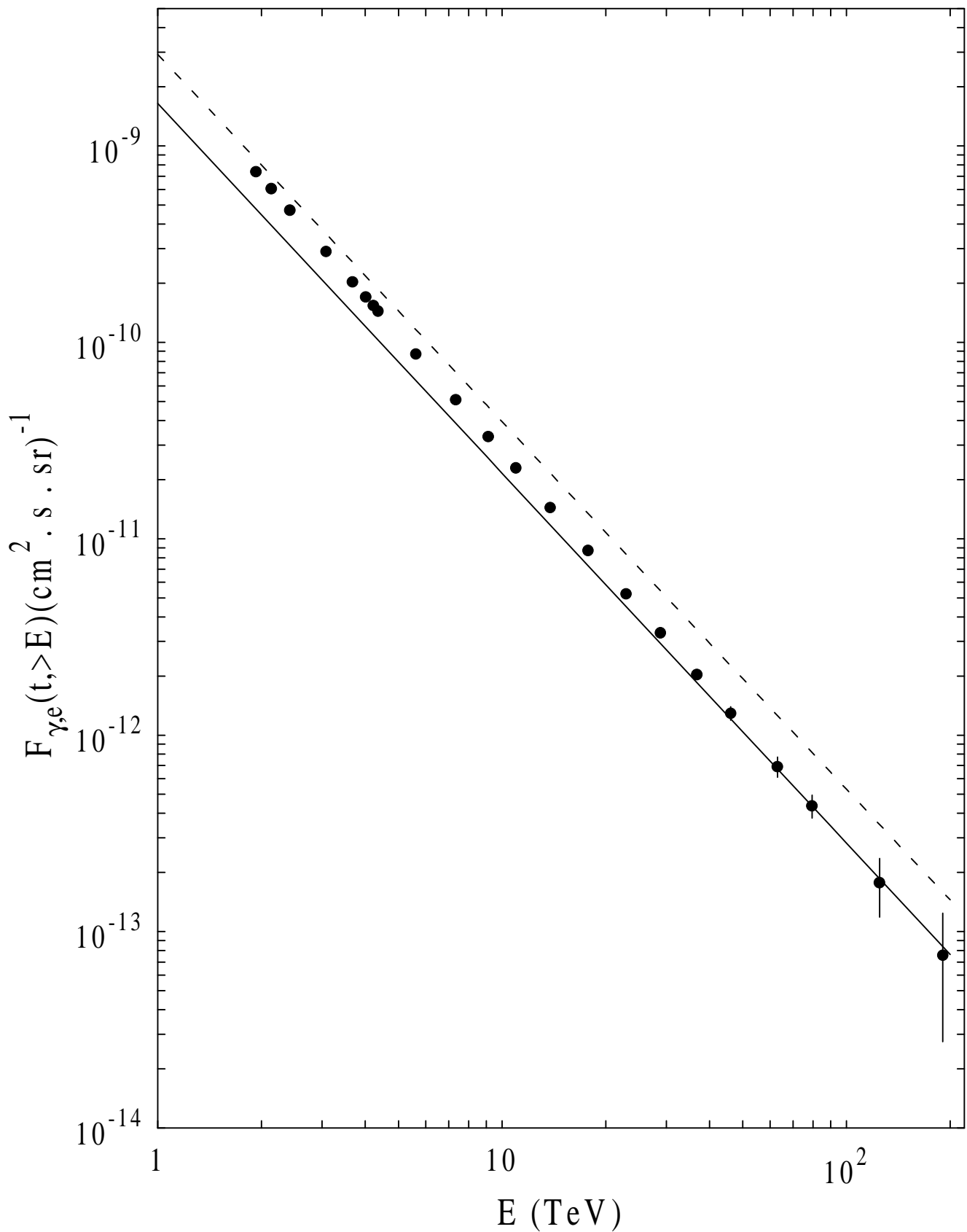
\end{document}